\documentclass{INTERSPEECH2024}
\usepackage{tikz}
\usepackage{graphicx}
\usepackage{subfig}
\usepackage{multirow}
\usepackage{algorithm}
\usepackage{algpseudocode}
\usepackage{multirow}
\usepackage{color,soul}
\usepackage{svg}
\usepackage{pifont}

\usepackage{ifthen}
\definecolor{darkgreen}{RGB}{11, 153, 56}

\definecolor{extradarkgreen}{RGB}{22, 199, 78}
\newcommand{\cmark}{\ding{51}}%
\newcommand{\xmark}{\ding{55}}%

\interspeechcameraready 

\title{Self-Train Before You Transcribe}
\name{Robert}{Flynn}
\name{Anton}{Ragni}

\address{Department of Computer Science, The University of Sheffield, United Kingdom}
\email{\{rjflynn2, a.ragni\}@sheffield.ac.uk}
\keywords{speech recognition, long-context, self-attention}

\begin{document}

\maketitle
 
\begin{abstract}
When there is a mismatch between the training and test domains, current speech recognition systems show significant performance degradation. Self-training methods, such as noisy student teacher training, can help address this and enable the adaptation of models under such domain shifts. However, self-training typically requires a collection of unlabelled target domain data. For settings where this is not practical, we investigate the benefit of performing noisy student teacher training on recordings in the test set as a test-time adaptation approach. Similarly to the dynamic evaluation approach in language modelling, this enables the transfer of information across utterance boundaries and functions as a method of domain adaptation. A range of in-domain and out-of-domain datasets are used for experiments demonstrating large relative gains of up to 32.2\%. Interestingly, our method showed larger gains than the typical self-training setup that utilises separate adaptation data.
\end{abstract}



\section{Introduction}
Speech recognition models are usually trained on a collection of speech and text data, then remain fixed at inference/test time when they are used to transcribe new speech. As a consequence, the model holds a fixed prior, formed from the training data, on the distribution of words and speech for every new utterance that is processed. This fixed prior becomes problematic as the change in distribution between the training data and the inference data becomes larger, resulting in reduced performance \cite{likhomanenko2020rethinking}. 


Prior work \cite{khurana2021unsupervised, higuchi2021momentum, meng2019domain, higuchi2022advancing} has shown that self-training methods, such as pseudo-labelling, \cite{lee2013pseudo} can be used to adapt models under domain shifts. This typically involves training an initial \textit{teacher} model on a source domain, which is then used to produce the target labels for a \textit{student} model on the target domain. While these methods do not require labelled data, they usually assume that a collection of unlabelled data from the target domain is available. Recent work on test time adaptation \cite{wang2020tent} (TTA) proposes to instead adapt the model solely at test time, without the need for a separate adaptation set. This is advantageous when the target domain is not known in advance, or unlabelled data is either not available or cannot be shared due to privacy concerns. Additionally, target domain data collection can be expensive, and may need to be repeated over time due to domain drift.


Pseudo-labelling can also be used as a TTA method for automatic speech recognition (ASR) \cite{lee2023awmc}. For this, the pseudo-labelling-based adaptation is performed solely on the current recording, prior to transcribing it. This process also allows for the transfer of information across utterance boundaries via gradient descent, relaxing the independence assumption that is usually made when segmenting a recording into separate utterances. We draw parallels to the dynamic evaluation approach \cite{mikolov2010recurrent, krause2018dynamic} that has proved effective in language modelling, where training models on the previous history enables them to better exploit recurring patterns, that occur outside the effective context window. Extensions to pseudo-labelling such as noisy student teacher (NST) training \cite{xie2020selfnst, park2020improved,  higuchi2021momentum}, where noise/augmentation is added to the inputs of the student model to make prediction more challenging, have proven effective for self-training approaches. Hence, in this work, we investigate the use of NST at test time as a method of TTA. A breakdown of our main contributions is given as follows:
\begin{enumerate}
    \item We propose ($\S$ \ref{sec:method}) a method that substantially improves over prior work, demonstrating that NST is more effective than standard pseudo-labelling for TTA. We find ($\S$ \ref{sec:transf}) that the use of augmentation is particularly crucial when the domain mismatch is large.
    \item We show ($\S$ \ref{sec:nstvsnsti}) that our method leads to better performance, while using $100\times$ less data, than a more standard self-training approach that uses separate adaptation data. 
    \item We show ($\S$ \ref{sec:rec_dur}) that the method improves with longer recordings and that the local context is most useful for the adaptation.

    
\end{enumerate}

\noindent The model checkpoint and code to reproduce our results are made available here\footnote{\url{www.github.com/robflynnyh/Self-Train-Before-You-Transcribe}}.

\section{Method}
\label{sec:method}

\begin{figure*}[t]
    \centering
   
    \begin{minipage}{.48\textwidth}
        \includegraphics[width=8.5cm]{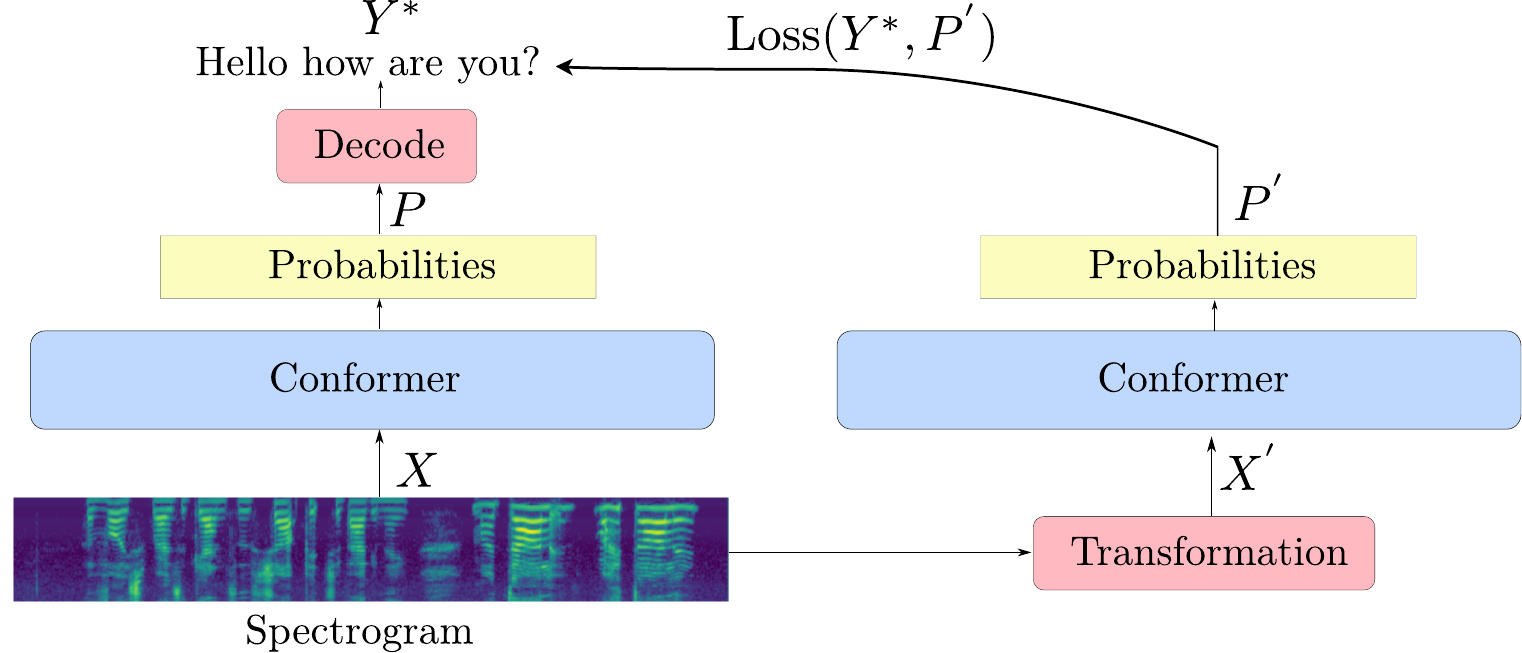}

    \caption{ Depiction of the NSTI method}
    \end{minipage}\hfill
    \begin{minipage}{.48\textwidth}
    \begin{algorithm}[H]
    \footnotesize
    \caption{Noisy Student Teacher at Inference (NSTI)}\label{alg:NSTI}
    \begin{algorithmic}[1]
    \Require{Recording $R$, model $M$, number of epochs $n$}
    \State{Segment $R$ into $S_1, S_2, \dots, S_m$ \label{alg:segment}}
    \State{Shuffle the segments $S_1, S_2, \dots, S_m$}
    \For{$i = 1$ to $n$}
        \For{each segment $S_j$}
            \State{$X = S_j$}
            \State{Apply Transfomation: $X^{'} = \mathrm{Transformation}(X)$}
            \State{Compute probabilities: $P = M(X)$, $P^{'} = M^{'}(X^{'})$}
            \State{Decode $P$ to obtain target labels: $Y^{*} = \mathrm{Decode}(P)$}
            \State{Obtain gradients using the loss between $P^{'}$ and $Y^{*}$}
            \State{Update $M$ and $M^{'}$ using the gradients}
        \EndFor
    \EndFor
    \State{Obtain predictions for recording $R$ using the updated model $M$} 
    \end{algorithmic}
    \end{algorithm}
    \end{minipage}

\end{figure*}

The following section presents our method of TTA, which we refer to as Noisy Student Teacher at Inference (NSTI). This is depicted in figure \ref{alg:NSTI}, with algorithm \ref{alg:NSTI} covering the process that is used to transcribe recordings with NSTI. For this work CTC \cite{graves2006connectionist} based acoustic models are used.

For NSTI 2 spectrograms $X$ and $X^{'}$, where $X^{'} = \mathrm{Transformation}(X)$, are used as inputs to a teacher model $M$ and a student model $M^{'}$ which are identical and always share the same parameters. The Transformation used to produce $X^{'}$ can be an augmentation strategy such as SpecAugment \cite{park2019specaugment}. Output probabilities $P = M(X)$ and $P^{'} = M^{'}(X^{'})$ are obtained from the models. The models are then updated based on the loss between $P^{'}$ and a decoded label sequence $Y^{*}$ that is produced from decoding $P$.
Practically, as the student and teacher models are identical, this method only requires one forward and backwards pass per step through a single model.

For transcribing recordings using our method, the following process is used: first, the recording is segmented. Self-training is then performed on each segment in the recording in a random order, and repeated for $n$ epochs. A final pass is then performed over the recording to obtain the model's predictions. Note that the method is performed over individual recordings and not the Dev/Test set as a whole.


 
\subsection{Transformations}
\label{sec:transforms}
For the transformation function, the frequency masking component of SpecAugment \cite{park2019specaugment} is used for our primary experiments. The use of time masking was not found to be beneficial. In $\S$ \ref{sec:transf} we also experiment with a range of other transformations, which are as follows: Identity transform, where no change is applied; Random noise, where Gaussian noise is added to the spectrogram; CutOut \cite{devries2017improvedcutout}, where $n$ rectangles are masked out from the spectrogram. 

\section{Prior Work}
\label{sec:priorwork}

Self-training methods can be used for the purpose of adaptation \cite{khurana2021unsupervised, higuchi2021momentum, meng2019domain, higuchi2022advancing}, these methods typically use a separate training set to adapt a model to a target domain. In contrast, TTA methods use only the data available at test/inference time, such as the current recording or utterance. Entropy minimisation (EM), originally presented as a semi-supervised learning approach \cite{grandvalet2004semiem}, has proven effective as a method of TTA for computer vision \cite{wang2020tent}  and ASR \cite{lin2022listen}. The Single-Utterance Test-time Adaptation (SUTA) \cite{lin2022listen} is an example of a TTA approach for ASR that uses an EM-based technique. This method adapts models on a single utterance and then discards the adapted model for the following utterance. While SUTA is appropriate in an $i.i.d$ scenario, recordings containing multiple utterances with highly correlated features are present at test time in most scenarios. Therefore, to improve performance our work leverages the entire recording, rather than a single utterance.

To benefit from updates from preceding utterances, AWMC \cite{lee2023awmc}, adopts a psuedo-labelling approach towards TTA. This method focuses on an online scenario where utterances are trained on and transcribed in their natural order. Due to problems with model collapse, AWMC uses 2 models to serve as the teacher model, with weights that are an exponential moving average (EMA) of the students. While \cite{lee2023awmc} investigates TTA for an online setting where only prior utterances are available, our work focuses on an offline approach that also uses future utterances for improved performance. Additionally, the use of augmentation is not explored in \cite{lee2023awmc}. In our work, we find that the use of an EMA teacher is not needed to prevent model collapse. This may be due to a more stable base model and the use of augmentation. Attempts to include an EMA teacher also resulted in reduced performance due to the teacher adapting at a slower rate, however in scenarios where model collapse does occur this would be beneficial.

Other work in natural language processing, also investigates a form of TTA \cite{kedia2021keeplearning} where pseudo-labelling is used at test-time to adapt the model via gradient descent. In this work, low-quality pseudo-labels are filtered based on confidence, and the model is trained to predict the pseudo-labels while maintaining similarity to the pre-trained weights. Other work, in language modelling presents a technique referred to as dynamic evaluation \cite{mikolov2010recurrent, krause2018dynamic}, where gradient descent is used to update the model based on the prior text history. This enables the model to exploit recurring patterns in the sequence, which occur outside the effective context window. However, it generally relies on the availability of a ground truth text history.

\section{Experimental Setup}

\subsection{Datasets}
For training the initial pre-adapted baseline model, the collection of Spotify podcasts totalling 58K hours provided in \cite{clifton2020spotify} is used. For the experiments in this work, a range of in-domain and out-of-domain datasets are included in order to properly evaluate our method, which are as follows:
\textbf{Earnings-22} \cite{del2022earnings}, which consists of earnings report meetings with a diverse range of accents. The train/dev/test splits from \cite{gandhi2022esb} are used in this work. Each meeting lasts up to 2 hours in length, with 5.5/5.6 in the dev and test sets respectively. \textbf{Rev16}, which is a collection of 30 podcast episodes, we use the 16 episodes reported on in \cite{radford2023robust}. Rev16 can be viewed as our in-domain test set due to its similarity to our training data. \textbf{Tedlium} \cite{hernandez2018ted}, which is a collection of 10-20 minute TED talks, typically involving a single speaker. \textbf{Chime6} \cite{watanabe2020chime}, which is used as a highly out-of-domain test set, we use only the first distant microphone array from this data, resulting in large amounts of background noise. Channels of the first array are combined by averaging the spectrograms prior to normalisation.

\subsection{Model Configuration}
The model uses a Conformer \cite{gulati2020conformer} based architecture, with the Fast Conformer subsampling configuration \cite{rekesh2023fast}. For our method, we find it essential to replace batch normalisation \cite{ioffe2015batch} with batch renormalization \cite{ioffe2017batchrenorm} in the conformer convolution modules, and do not update the batch statistics during the self-training process. A similar approach is taken in \cite{higuchi2022advancing}, where group normalisation is used, however, we find this to be unstable during the initial training stage. The model consists of 6 layers with roughly 90 million parameters. Training and adaptation is performed with a context window/segment length of 162 seconds.


\subsection{Hyperparameters}
\label{sec:hyperparams}
All hyperparameters (i.e. augmentation, epochs, learning rate) are tuned using a random search on the development set of Tedlium for our primary experiements. Dataset-specific tuning is also explored in $\S$ \ref{sec:hyperparamdepend}. All experiments use a batch size of 2, composed of $2$ copies of a single segment/utterance. Madgrad \cite{defazio2022adaptivitymadgrad} is used as the optimizer. For our primary results that use SpecAugment the following hyperparameters are used: $6$ frequency masks with a maximum size of $34$, $5$ epochs and a learning rate of $9e-5$. 

For segmenting recordings the sliding window scheme described in \cite{flynn2023much} is used. Specifically, the recording is chunked into segments/windows equal to the model's context window using a moving window with a stride of $12.5\%$ of the segment length. When transcribing the recording, probabilities from segments that overlap are averaged to obtain the final prediction. However, this process is not essential to the method, and standard utterance boundaries can also be used. 

\section{Experimental Results}
All evaluations reported in the tables are repeated 3 times, with mean statistics reported. We find the variation between repeats to be low, with a standard deviation of around $0.01-0.1$. The following subsections will break down and discuss our findings.
\label{sec:results}

\vspace{-1em}
\subsection{How effective is the method?}

\vspace{-1em}
\begin{table}[hbt]
    \centering
    \caption{WERs (Dev/Test) when using various training settings. *Method from prior work \cite{lee2023awmc}.}
    
    \scriptsize
    \begin{tabular}{c|c|c|c|c|c}
      Setting   & Aug &  TED & E-22 & Chime6 & Rev16  \\\hline
        Shuffled  & \cmark & \textbf{6.6/5.8}& \textbf{18.9/14.9} & \textbf{56.6/59.4} & \textbf{14.2}\\
        Ordered & \cmark &\textbf{ 6.6}/5.9 & 19.5/15.4 & 57.3/61.8 & \textbf{14.2} \\
        Online   & \cmark& 6.9/6.1 &  21.7/16.7  & 59.7/64.7 & \textbf{14.2} \\
        AWMC* & \xmark & 7.0/6.2 & 23.4/18.1 & \textcolor{red}{88.9}/85.3 & \textcolor{red}{15.2} \\
        AWMC &  \cmark & 6.8/6.0 & 20.7/15.7 & 70.7/75.9 & \textbf{14.2}
        \\\hline
        Unadapted model & N/A &7.1/6.2 & 23.9/18.3 & 83.5/86.5 & 14.5
    \end{tabular}

    \label{tab:NSTISettings}
\end{table}
\vspace{-2em}
\begin{table}[hbt]
    \centering
    \caption{Real time factor of each setting}
    \scriptsize
    \begin{tabular}{c|c|c|c}
        Setting  & Aug & 1 Epoch & Total \\\hline
       Shuffled  & \cmark & 0.027 & 0.115 \\
       Ordered  & \cmark & 0.027  & 0.115 \\
       Online  & \cmark & 0.023 & 0.023 \\
       AWMC  & \xmark & 0.026 & 0.026\\
       AWMC  & \cmark & 0.026 & 0.097 \\\hline
        Unadapted model & N/A & N/A & 0.004 \\
        
    \end{tabular}
    
    \label{tab:rtf}
\end{table}

\noindent Table \ref{tab:NSTISettings} presents the results for our method using various settings. Settings with augmentation (Aug), use the frequency masking component of SpecAugment \cite{park2019specaugment} The real-time factors (RTF) for each setting are provided in table \ref{tab:rtf}. \textbf{Bold} text denotes the best results, \textcolor{red}{red} text denotes that the model performs worse after test-time adaptation.

The primary method described in algorithm \ref{alg:NSTI} is referred to as the Shuffled setting. The results for this setting demonstrate substantial gains over the unadapted model with word error reductions (WERRs) of 6.5\%, 18.6\%, 31.3\% and 2.0\% for Tedlium, Earnings-22 (E-22), Chime6 and Rev16 respectively. Note that our un-adapted baseline already shows competitive performance \cite{hernandez2018ted, gandhi2022esb, likhomanenko2020rethinking, radford2023robust} due to the large and diverse training dataset. Whereas prior TTA work in ASR \cite{lee2023awmc, lin2022listen} uses a baseline trained on Librispeech \cite{panayotov2015librispeech}, which consequently shows unrealistic gains from adaptation due to the narrow training domain. 

In general, the datasets that are the most different from the training domain see the largest gains. Recordings in Tedlium are also shorter which results in less data for adaptation and therefore lower gains (see $\S$ \ref{sec:rec_dur}). Results for Chime6 show that the method remains effective and stable even when the domain mismatch is very large and transcription quality is extremely poor.  Rev16, our in-domain test set shows the smallest gains. We hypothesise that this is due to the similarity of this data to our large training dataset.

\vspace{-1em}
\begin{figure}[hbt]
    \centering
    \includegraphics[width=6cm]{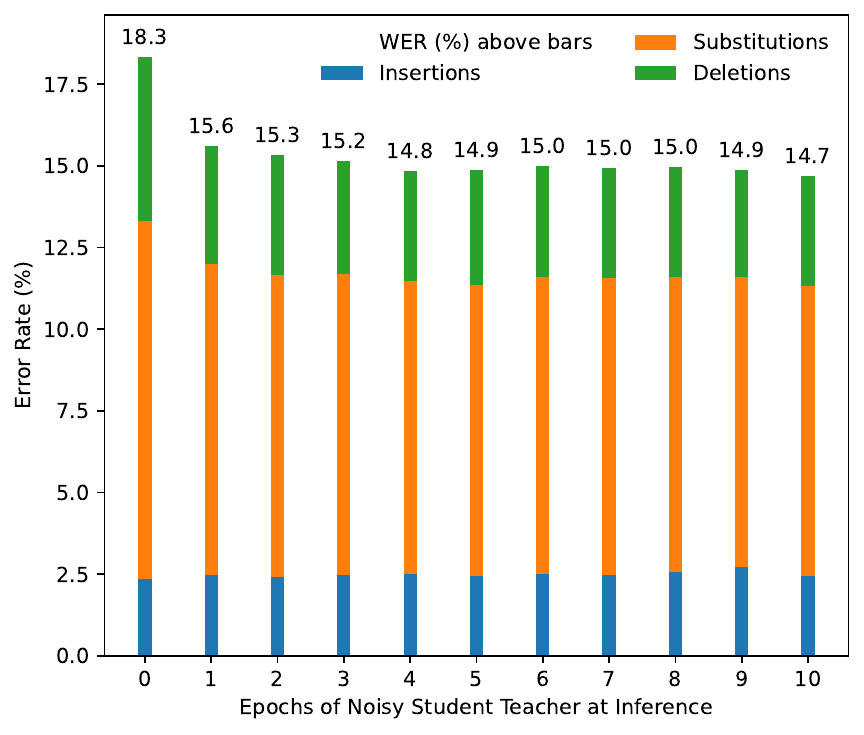}
    \caption{Error rates after each epoch of NSTI on Earnings-22}
    \label{fig:epocherrorrates}
\end{figure}
\vspace{-1em}
Word error rate (WERs), substitution, insertions and deletion statistics after each epoch of NSTI on Earnings-22 (test) are provided in figure \ref{fig:epocherrorrates}. We find that the majority of the benefit from the method is attained after the first epoch, with a WERR of 14.8\%. Therefore, adaptation can be stopped here if RTF is an issue. The method benefits from a reduction in the number of substitutions and deletions. Insertions see a small increase of 4.4\% by epoch 5, with deletions showing the largest reduction of 30.3\%. On the Chime6 dataset, the model is initially unconfident, with a very high deletion rate of 83.8\%, therefore the method is very beneficial for this data, with the deletion rate reducing to 37.2\% after NSTI. 

While shuffling the data during NSTI is helpful, results from the Ordered setting demonstrate that it is not necessary for good performance. We also present an online setting where utterances are processed in their natural order, and the final predictions are taken from the teacher model at each time-step. This shows worse performance as it can not leverage future utterances, and only runs for one epoch. The AWMC method \cite{lee2023awmc} shows a small WERR of 2.1\%/1.1\% on Earning-22, with the model degrading on Chime6 and Rev16. This is primarily due to not incorporating augmentation, as including augmentation causes the results to improve. As AWMC is an online method the use of this approach with augmentation may be preferable to the online variation of NSTI. However, the WER is higher than other settings on Chime6, which may be due to the EMA teacher model updating at a slower rate.

\subsection{Dependency on hyperparameters}
\label{sec:hyperparamdepend}
The hyperparameters used in the results are tuned on the Tedlium development split. We also experiment with dataset-specific tuning. On Earnings-22 there was no change in results when tuning was performed on the datasets Dev set. For Chime6 there was a small further WERR of 2.3\%/4.0\%. This demonstrates that the method is not overly sensitive to hyperparameter choice, and does not require repeated re-tuning when the test domain changes. Which is important for scenarios where NSTI is likely to be useful.

\begin{figure}[h!]
    \centering
    \includegraphics[width=6cm]{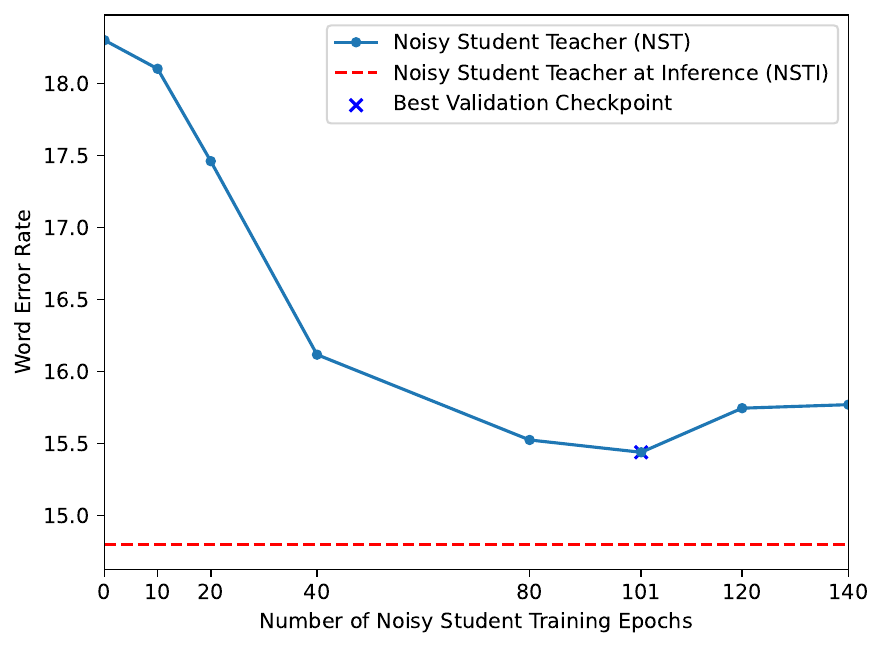}
    \caption{Comparison between NST and NSTI on Earnings-22}
    \label{fig:NSTvsNSTI}
\end{figure}



\vspace{-1em}
\subsection{Comparison to self-training on a separate training set}
\label{sec:nstvsnsti}
We provide comparisons to an NST self-training approach that uses the 105-hour Earnings-22 training set to adapt the model in figure \ref{fig:NSTvsNSTI}. For the NST training we use the method described in \cite{higuchi2021momentum}. Results show that our method (NSTI) improves over NST by 4\% when selecting the best checkpoint on the validation data. During NSTI only the current recording with a duration of around 1 hour is used, hence our method required $100\times$ less data. Performing NSTI after adapting the model on the training set did not result in any further gains. We believe this is due to the sharpening of the model's class distribution after NST training.

Self-training on separate adaptation data may still be preferred in scenarios where the RTF is an issue and the target domain is known and constant. This is because NSTI needs to be performed on every recording. For example, adapting on a recording from Earnings-22 and then evaluating on other recordings from the same test set led to a WERR of -3.9\% after the first epoch, and -17.4\% after the fifth. 

\subsection{Comparison of transformation functions}
\label{sec:transf}
\vspace{-1em}
\begin{table}[hbt]
    \centering
    \caption{WERs when using various transformation functions.}
    \scriptsize
    \begin{tabular}{c|c|c|c|c}
      Transform   &  TED & E-22 & Chime6 & Rev16  \\\hline
        SpecAugment  & 6.6/\textbf{5.8} & \textbf{18.9}/14.9 & 56.6/59.4 & \textbf{14.2}\\
        Identity & 7.0/6.1 & 22.6/17.4 & \textcolor{red}{100.0/100.0} & \textcolor{red}{15.0} \\

        Noise & 6.6/5.9  & \textcolor{red}{24.2/19.2} & \textcolor{red}{99.6/97.3} & \textcolor{red}{18.5} \\
        CutOut & \textbf{6.4}/\textbf{5.8} & \textbf{18.9}/\textbf{14.5} & \textbf{54.9}/\textbf{56.9} & \textcolor{red}{37.9} \\
        \hline
        Unadapted model & 7.1/6.2 & 23.9/18.3 & 83.5/86.5 & 14.5
    \end{tabular}

    \label{tab:transfns}
\end{table}
\noindent A comparison of the different transformation functions described in $\S$ \ref{sec:transforms} is provided in table \ref{tab:transfns}. The identity transform (no transformation) shows the worst performance, with small improvements on Tedlium/Earnings-22 and worse performance than the unadapted model on Chime6/Rev16. This explains the poor performance of AWMC \cite{lee2023awmc}, which uses this transformation. We find that the use of augmentation is especially crucial for datasets, such as Chime6, where the deletion rate is high, otherwise, the model learns to only output blank tokens.

Using random noise as the transformation function resulted in degradation on all datasets other than Tedlium. Cutout \cite{devries2017improvedcutout} showed the best performance on all datasets other than Rev16, where the model performed much worse than the unadapted model. The degradation seen on various datasets for the Identity, Noise, and Cutout transforms is because these methods were more sensitive to hyperparameter choices (which were tuned on Tedlium). Overall, the frequency masking component of SpecAugment showed the most consistent performance and outperformed the unadapted model by a large margin on all datasets.

\subsection{Impact of recording duration}
\vspace{-1em}
\label{sec:rec_dur}
\begin{figure}[hbt]
    \centering
    \includegraphics[width=6cm]{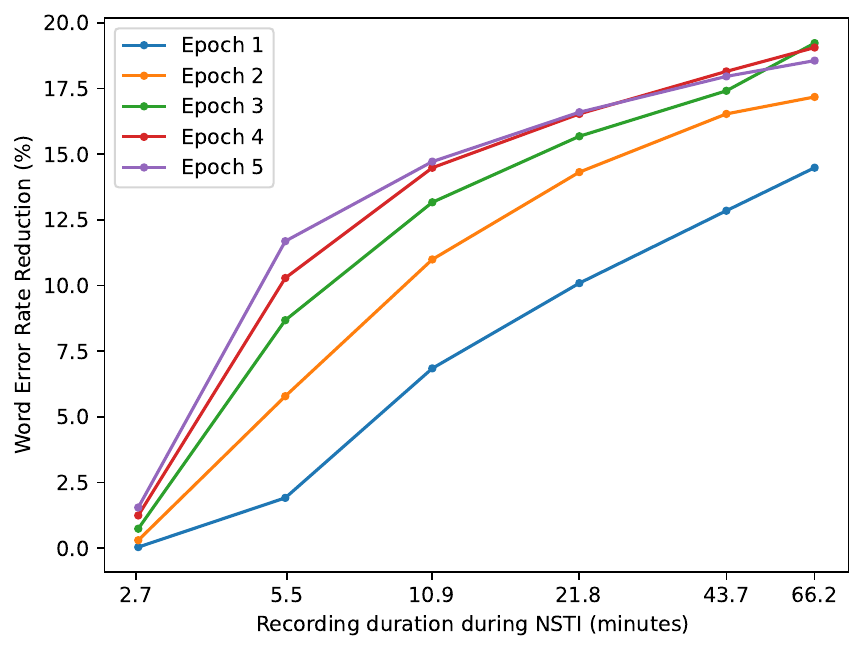}
    \caption{WER as the recording duration for NSTI is increased}
    \label{fig:recdur}
\end{figure}

\noindent Figure \ref{fig:recdur} presents results analysing the effect of recording duration on WERR. For this experiment, all recordings from Earnings-22 Dev/Test that are 1 hour or greater in duration are used. Each recording is partitioned into segments of 2.7, 5.5, 10.9, 21.8 and 43.7 minutes using the moving window scheme discussed in $\S$ \ref{sec:hyperparams}. NSTI is then performed separately on each of these partitions. Results when using the entire recording, which range from 61-76 minutes ($\mu = 66.2$), are also provided.

When the recording duration is equal to the context window of the model (2.7 minutes) we see very small gains of less than 1\%. As the model is already attending over this data it is likely that some form of adaptation is already being performed. As expected the local context beyond the model's context window is more valuable, for example performing two epochs with a recording duration of 10.9 minutes is more beneficial than 1 epoch at 21.8 minutes. This helps explain the discrepancy in performance we see in $\S$ \ref{sec:nstvsnsti} between NSTI and NST. Self-training on separate adaptation data requires much more data because this data is much less representative of any recording at test time.




\vspace{-1em}
\section{Conclusion}
ASR models often degrade considerably in performance when the domain mismatch between training and testing/inference is large. In this work, to help address these challenges, we propose a self-training approach, which applies the noisy student teacher training framework on recordings at test time before transcribing them. Our proposed method improves considerably over prior test time adapatation methods for ASR. Additionally, the results demonstrate that this work leads to better performance than a more typical approach which uses a separate 105 hour training set for adaptation. We find that this is due to the high corellation between a current utterance and surrounding utterances in a recording. While, the method is very effective, it does not take into account the sequential nature of utterances in a recording, we plan to investigate this, alongside other augmentation strategies in our future work.

\section{Acknowledgements}
This work was supported by the CDT in Speech and Language Technologies (SLT) and their Applications funded by UKRI [grant number
EP/S023062/1].

\bibliographystyle{IEEEtran}
\bibliography{mybib}

\end{document}